\documentstyle[12pt,epsfig]{article} 

\def\NPB#1#2#3{Nucl. Phys. {\bf B#1}, #3 (19#2)}
\def\NPPS#1#2#3{Nucl. Phys. Proc. Suppl. {\bf #1}, #3 (19#2)}
\def\PLB#1#2#3{Phys. Lett. {\bf B#1}, #3 (19#2)}

\def\PRD#1#2#3{Phys. Rev. {\bf D#1}, #3 (19#2)}
\def\PRL#1#2#3{Phys. Rev. Lett. {\bf#1}, #3 (19#2)}

\def\ZPC#1#2#3{Zeit. f\"ur Physik {\bf C#1}, #3 (19#2)}

\newcommand{\postscript}[2]
{\setlength{\epsfxsize}{#2\hsize}
\centerline{\epsfbox{#1}}}
\newcommand{\agt}{ \mathop{}_{\textstyle \sim}^{\textstyle >} }
\newcommand{\alt}{ \mathop{}_{\textstyle \sim}^{\textstyle <} }
\newcommand{\mheavy}{m_{\rm heavy}}
\newcommand{\mlight}{m_{\rm light}}
\newcommand{\tev}{\,{\rm TeV}}
\newcommand{\tr}{{\rm Tr}\,}

\begin{document}

\begin{titlepage}
\begin{flushright}
{\rm
IASSNS--HEP--98--85\\
LBNL--42428\\
RU--97--98\\
October 1998\\
hep-ph/9810500
}
\end{flushright}
\vskip 1cm

\begin{center}
{\Large \bf
Solving the Supersymmetric Flavor Problem with \\
Radiatively Generated Mass Hierarchies
}
\vskip 0.8cm
{\large
Jonathan L. Feng,${}^1$ Christopher Kolda,${}^2$ and
Nir Polonsky${}^3$\\}
\vskip 0.4cm
{\small\em ${}^1$ School of Natural Sciences,
Institute for Advanced Study\\
Princeton, NJ 08540, USA\\
${}^2$ Theoretical Physics Group, Lawrence Berkeley National
Laboratory\\
University of California, Berkeley, CA 94720, USA\\
${}^3$ Department of Physics and Astronomy, Rutgers University\\
Piscataway, NJ 08854, USA\\
}
\end{center}
\vskip .8cm


\begin{abstract}
The supersymmetric flavor problem may be solved if the first and
second generation scalars are heavy (with multi-TeV masses) and
scalars with large Higgs couplings are light (with sub-TeV masses).
We show that such an inverted spectrum may be generated radiatively;
that is, from initial conditions where all scalar masses are multi-TeV
at some high scale, those with large Higgs couplings may be driven
asymptotically to the weak scale in the infra-red.  The lightness of
third generation scalars is therefore a direct consequence of the
heaviness of third generation fermions, and fine-tuning is avoided
even though the fundamental scale of the soft supersymmetry breaking
parameters is multi-TeV.  We investigate this possibility in the
framework of the usual Yukawa quasi-fixed point solutions.  The
required high scale boundary conditions are found to be simple and
highly predictive.  This scenario also alleviates the supersymmetric
$CP$ and Polonyi problems.
\end{abstract}

\setcounter{page}{1}

\end{titlepage}
\setcounter{footnote}{0}


\section{Introduction}

If low-energy supersymmetry (SUSY) is realized in nature, the
effective Lagrangian must contain many new mass parameters that
explicitly, but softly, break supersymmetry.  The requirement that
large quadratic divergences not be reintroduced in the electroweak
breaking sector is often taken to suggest that these soft
SUSY-breaking (SSB) parameters are at the scale $\mlight \alt 1\tev$.
On the other hand, stringent flavor changing constraints require that
many of the soft scalar masses either be at the scale $\mheavy \sim
10\tev$ or fall into highly constrained patterns~\cite{constraints}.
The tension between these requirements is the supersymmetric flavor
problem.

This problem may be resolved, however, if the scalars have an inverted
mass hierarchy relative to the fermions~\cite{dgpt,dp,ckn}.  In such a
scenario, the scalars of the first two generations are at the scale
$\mheavy$.  This highly suppresses supersymmetric contributions to
flavor (and $CP$) violation involving the first two families, where
the constraints are most stringent.  At the same time, the scalar
partners of the heavy fermions, which interact through large Yukawa
couplings with the Higgs bosons, are at the scale $\mlight$, avoiding
fine-tuning in the Higgs sector.  Note that, because the scalars of
the first two generations interact very weakly with the Higgs bosons,
they may be significantly heavier without destabilizing the gauge
hierarchy.

This inverted hierarchy of scalar masses has been analyzed in a number
of studies, and it has been argued that it may be created by dynamical
mechanisms at high~\cite{dp,U1} or intermediate~\cite{horizontal}
energies.  The experimental signatures of such scenarios have also
been studied.  Observable effects of the light supersymmetric
particles have been considered in Refs.~\cite{an,bphys}, and the
non-decoupling effects of very massive superparticles have been
discussed in Refs.~\cite{so,kn}.

In this paper, we note that there is no {\it a priori\/} need to
impose this hierarchy among the scalar masses at some high scale, such
as the grand unified theory (GUT) or Planck scale, in order to realize
the hierarchy at the weak scale. Instead, we demonstrate that even if
{\em all} soft scalar masses have multi-TeV values at some high scale
boundary, the mass hierarchy may be generated radiatively.  In this
scheme, for specific ratios of the SSB parameters which we will
determine, the third generation scalars are driven to the light scale
by large Yukawa couplings.  The lightness of third generation scalars
and heaviness of third generation fermions are therefore intimately
connected, and fine-tuning is avoided, even though the fundamental
scale of the SSB parameters (the gravitino mass) is $\sim 10$ TeV.

We will demonstrate this idea in the context of scenarios in which
large Yukawa couplings saturate their infra-red quasi-fixed points
(QFPs).  In this case, the relevant SSB parameters will be seen to
have simultaneous (approximate) zero fixed-points.  The required
boundary conditions for such fixed points to exist will be seen to be
remarkably simple and highly predictive, though also highly
constrained~\cite{susy}.\footnote{In a related, but orthogonal,
approach to the supersymmetric flavor problem, one may search for
models in which scalar mass degeneracy, as opposed to a scalar mass
hierarchy, is generated by fixed points~\cite{otherfp}.}

\section{Inverted Hierarchy Models}
\label{sec:invhier}

We first review the constraints on supersymmetric models with inverted
scalar mass hierarchies.

A supersymmetric scenario is fine-tuned if there are large
cancellations in the conditions for electroweak symmetry breaking:
\begin{eqnarray}
\frac{1}{2} m_Z^2 &=& \frac{m_{H_d}^2 - m_{H_u}^2 \tan^2\beta }
{\tan^2\beta -1} - \mu^2, \nonumber \\
2 m_3^2 &=& (m_{H_u}^2 + m_{H_d}^2 + 2 \mu^2)\sin2\beta \ .
\label{finetune}
\end{eqnarray}
In these equations, $m_{H_u}$ and $m_{H_d}$ are the SSB Higgs boson
masses, $m_3^2$ is the soft bilinear scalar coupling of the two Higgs
doublets, $\mu$ is the Higgsino mass parameter, and $\tan\beta =
\langle H_u^0 \rangle /\langle H_d^0\rangle$ is the usual ratio of
Higgs vacuum expectation values.  In models with hierarchical scalar
masses, these conditions have a number of implications resulting from
the fact that the light-heavy scalar mass hierarchy and, hence, the
Higgs parameters of Eq.~(\ref{finetune}) are not stable against
radiative corrections~\cite{drees,dgpt}.

Several of these implications are evident even at
one-loop~\cite{drees}.  For the soft scalar mass parameter $m_i$, the
one-loop renormalization group equation is
\begin{eqnarray}
\left. \frac{d\,m^{2}_{i}}{d t} \right|_{\rm 1-loop} &=&
4\sum_{a}C_{a}(i)\alpha_{a}|M_a|^{2}
- \frac{1}{4\pi}\sum_{pq}h_{ipq}^{2}A_{ipq}^{2}  \nonumber \\
&&- \frac{1}{2}{Y_{i}}\alpha_{Y} \sum_{j} Y_{j} m_{j}^{2}
- \frac{1}{4\pi}\sum_{pq}h_{ipq}^{2}\left(m_{i}^{2}+
m_{p}^{2}+m_{q}^{2}\right) \ ,
\label{rge}
\end{eqnarray}
where $t = \ln (M_X^2 / Q^2) / 4\pi$, and $M_X$ is the high scale
boundary. The index $a$ runs over gauge groups, $C_a(i)$ are quadratic
Casimir invariants,\footnote{For the U(1) gauge group, $C_{a} = Y^2$
for scalars with hypercharge $Y$, and for the SU(N) gauge groups, $C_a
= (N^2-1)/2N$ for scalars in the fundamental representation.} and $Y$
denotes hypercharge.  $M$, $h$, and $hA$ are gaugino masses, Yukawa
couplings, and trilinear scalar couplings, respectively. Summations
over scalar indices implicitly include summations over color and weak
isospin.  In general, of course, the masses need not be
flavor diagonal, and one must evolve a general mass matrix.
Discussion of the off-diagonal masses and their constraints will be
deferred to Sec.~\ref{sec:summary}.

Each of the four terms of Eq.~(\ref{rge}) leads to a constraint for
generating and maintaining a scalar mass hierarchy. From the first and
second terms, we see that gauginos masses and trilinear scalar
couplings must be at the light scale. From the third term, which
arises from quartic scalar gauge interactions, it is evident that the
hypercharge trace must roughly satisfy $\sum_j Y_{j}m_{j}^{2} \alt
\mlight^2$. From the fourth, one-loop corrections to light scale
masses of the form $h^2 \mheavy^2$ lead to the approximate upper bound
$\mheavy \alt \mlight / h$.

Even if the three constraints and upper bound mentioned above are
satisfied, two-loop gauge interactions threaten to drive the light
scalar masses negative.  These two-loop corrections are given
by~\cite{rge}:
\begin{equation}
\left. \frac{d\,m^{2}_{i}}{d t} \right|_{\rm 2-loop} \propto
\sum_a \sum_j - t_{a}(j)C_{a}(i)\alpha_{a}^{2}m_{j}^{2} \ ,
\label{TL}
\end{equation}
where $a$ again sums over all gauge groups, $t_{a} = Y^2$ for
hypercharge, and $t_{a} = \frac{1}{2}$ for fundamentals of SU(N).  To
avoid tachyonic states and color-breaking minima, these must be
compensated by positive contributions from gaugino
masses~\cite{dgpt,am,ag}. This observation leads to lower bounds on
gaugino masses that are most stringent in models with high-energy
mediation of the heavy SSB parameters, where the evolution interval
(the logarithm) is maximized.

\section{Radiative Hierarchy with Low $\tan\beta$}

Now we present a first concrete example of the generation of an
inverted hierarchy through renormalization group evolution.  This will
serve as a simple illustration of the idea. A more complicated, but
more satisfactory, scenario will be discussed in the following
section.

We will consider the minimal supersymmetric standard model (MSSM) with
superpotential
\begin{equation}
W = h_u^{ij} H_u Q_i U_j + h_d^{ij} H_u Q_i D_j +
h_e^{ij} H_d L_i E_j  +\mu H_uH_d\ ,
\end{equation}
where $H_u$ and $H_d$ are the up- and down-type Higgs superfields, $Q$
and $L$ are the quark and lepton doublets, $U$, $D$, and $E$ are the
up-type quark, down-type quark, and charged lepton singlets,
respectively, and the indices $i$ and $j$ denote generations.

We begin by considering the case of low $\tan\beta$.  In this
scenario, the only significant Yukawa coupling is the top quark Yukawa
$h_t \equiv h_u^{33}$.  As noted above, the stability of light-heavy
scalar hierarchies requires gaugino masses and trilinear scalar
couplings to be at the light scale.  Scenarios in which this arises
naturally will be described in Sec.~\ref{sec:Rsymm}.  Assuming this to
be true, and further neglecting the Tr$[Ym^2]$ term, we find that the
scalar masses renormalized by the top Yukawa satisfy
\begin{equation}
\frac{d\,{ \bf m^2}}{d t} = 
\frac{h_t^2}{4\pi} {\bf X}_{\rm low} {\bf m^2} \ ,
\end{equation}
where
\begin{equation}
{\bf X}_{\rm low} = - \left(
\begin{array}{ccc}
3 & 3 & 3 \\
2 & 2 & 2 \\
1 & 1 & 1 \end{array} \right)
\end{equation}
and ${\bf m^2} = (m_{H_u}^2, m_{U_3}^2, m_{Q_3}^2)^T$.  Two
eigenvectors of ${\bf X}_{\rm low}$ have eigenvalue 0; the third,
${\bf\hat{m}^2} = (3,2,1)^T$, has eigenvalue $-6$.  Arbitrary boundary
conditions may be evolved by first decomposing them along the three
eigenvectors~\cite{baggerwise}. The components parallel to the
eigenvectors with zero eigenvalue are constants of the evolution, and
the component parallel to ${\bf\hat{m}^2}$ is asymptotically damped to
zero.  If the initial conditions are dominated by their
${\bf\hat{m}^2}$ component, the three scalar masses will have, subject
to the assumptions above, simultaneous fixed points at zero, thereby
creating a scalar mass hierarchy.  Large components along the other
two eigenvectors are not allowed as they lead either to tachyons or
large and negative $m_{H_u}^2$.

To determine whether the fixed points for the mass parameters are
reached rapidly enough, let us consider scenarios in which the top
quark Yukawa is near its quasi-fixed point (QFP)~\cite{qfp}.  In the
low $\tan\beta$ QFP scenario, $h_t$ is drawn to its QFP value of
$h_t^{\rm FP} \approx 1.1$ in the infra-red, irrespective of its value
at the high scale, as long as this value is not too small.  Weak scale
parameters are therefore insensitive to the exact value of the top
Yukawa at the high scale, which is attractive because our scenario may
then be realized without postulating specific and possibly complicated
relations between the parameters of the Yukawa and SSB sectors.  Given
the relation $h_{t}(m_{t}) \simeq \left(0.95/ \sin\beta \right)\left(
m_t^{\rm pole}/{\rm 175 \, GeV}\right)$, we find $\tan\beta_{\rm FP}
\approx 1.8$ for the low $\tan\beta$ QFP scenario.\footnote{We ignore
here various subtleties associated with the value of the strong
coupling and with finite superpartner radiative corrections to
$m_{t}^{\rm pole}$.  These can lead to substantial corrections, but
may be absorbed in the relevant value of $\tan\beta_{\rm
FP}$~\cite{fixed}.}  Such low values of $\tan\beta$ are currently
probed in Higgs boson searches.

In the QFP scenario, it is possible to solve analytically for the low
energy values of the soft scalar masses in terms of the high scale
boundary conditions, which we denote by zeroes~\cite{solution}:
\begin{eqnarray}
m_{H_u}^2 &\simeq& m_{H_u}^{2}(0) + 0.52 M_{1/2}^2 - 3 \Delta m^2
  \nonumber \\
m_{H_d}^2 &\simeq& m_{H_d}^{2}(0) + 0.52 M_{1/2}^2 \nonumber \\
m_{Q_i}^2 &\simeq& m_{Q_i}^{2}(0) + 7.2 M_{1/2}^2 - \delta_i
  \Delta m^2 \nonumber \\
m_{U_i}^2 &\simeq& m_{U_i}^{2}(0) + 6.7 M_{1/2}^2 - \delta_i 2
  \Delta m^2 \label{lowtanbeta} \\
m_{D_i}^2 &\simeq& m_{D_i}^{2}(0) + 6.7 M_{1/2}^2 \nonumber \\
m_{L_i}^2 &\simeq& m_{L_i}^{2}(0) + 0.52 M_{1/2}^2 \nonumber \\
m_{E_i}^2 &\simeq& m_{E_i}^{2}(0) + 0.15 M_{1/2}^2 \nonumber \ ,
\end{eqnarray}
where
\begin{eqnarray}
\Delta m^2 &\simeq& \frac{1}{6} \left[ m_{H_u}^{2}(0) +
m_{Q_3}^{2}(0) + m_{U_3}^{2}(0) \right] r \nonumber \\
 && + M_{1/2}^2 \left( \frac{7}{3} r - r^2
\right) + \frac{1}{3} A_0 \left(\frac{1}{2} A_0 - 2.3 M_{1/2}
\right) r \left( 1- r \right) \ ,
\label{deltam}
\end{eqnarray}
and, for simplicity, we have assumed a common gaugino mass $M_{1/2}$
and trilinear scalar coupling $A_0$ at the high scale, which is
identified with the scale of coupling constant unification. The
subscript $i$ is a generational index; $\delta_1 = \delta_2 = 0$ and
$\delta_3 = 1$.  Finally, the parameter $r = \left[ h_t/h_t^{\rm FP}
\right] ^2 \leq 1$ is a measure of the proximity of the top Yukawa
coupling to its QFP value at the weak scale.

{}From Eq.~(\ref{lowtanbeta}) we see that the large Yukawa coupling
$h_t$ gives a large negative correction to the $H_u$, $Q_3$ and $U_3$
scalar masses. It is easy to verify that in the limit of $r \to 1$ and
neglecting $M_{1/2}$, the equation $m_{H_u} = m_{Q_3} = m_{U_3} = 0$
is solved by the boundary conditions
\begin{equation}
\left[ m_{H_u}^2 (0), m_{U_3}^2 (0), m_{Q_3}^{2}(0) \right]
= m_0^2 \, \left[ 3,2,1 \right] \ ,
\label{bc}
\end{equation}
as expected.  (This relation was also noted in Ref.~\cite{carena}.)
That is, even if all scalar masses are at some heavy scale $\mheavy
\sim 10\tev$, if the constraints of Eq.~(\ref{bc}) are satisfied, then
$m_{H_u}$, $m_{Q_3}$, and $m_{U_3}$ are still only $\sim\mlight$ in
the infra-red.  {}From the form of Eqs.~(\ref{lowtanbeta}) and
(\ref{deltam}), we see that these conclusions hold, roughly, as long
as $1-r \alt (\mlight/\mheavy)^2$ and deviations from the boundary
conditions of Eq.~(\ref{bc}) satisfy $\Delta m_{H_u}^2, \Delta
m_{Q_3}^2, \Delta m_{U_3}^2 \alt \mlight^2$.

The light-heavy hierarchy is, of course, also subject to the
constraints discussed in Sec.~\ref{sec:invhier}.  {}From
Eq.~(\ref{deltam}), we see that we require $M_{1/2}, A_0 \sim
\mlight$.\footnote{In fact, $A_0 \sim \mlight$ is not required in this
example if $r\to 1$; this will not hold in general, however.}  In
addition, the boundary conditions for $H_d$ and the other sfermions
are constrained by the requirement Tr$[Y m^2] \alt \mlight^2$; simple
boundary conditions, such as the condition that all of these other
scalar masses equal $m_0$, may be found to satisfy this constraint.
Finally, the zero fixed points of the mass parameters receive the
usual two-loop gauge corrections of Eq.~(\ref{TL}).  Because of large
group theoretical factors, the two-loop corrections to the light
sfermion masses are always more important than the one-loop Yukawa
correction. As noted above, for the light scalar squared masses to
remain positive, the negative two-loop corrections above must be
compensated by positive gaugino mass contributions.  The requirement
that there be no tachyons or color-breaking minima demands roughly
that $M_{1/2} \agt \sqrt{ \alpha_{s}/ 4\pi}\, m_{0}$.

\begin{figure}[ht]
\postscript{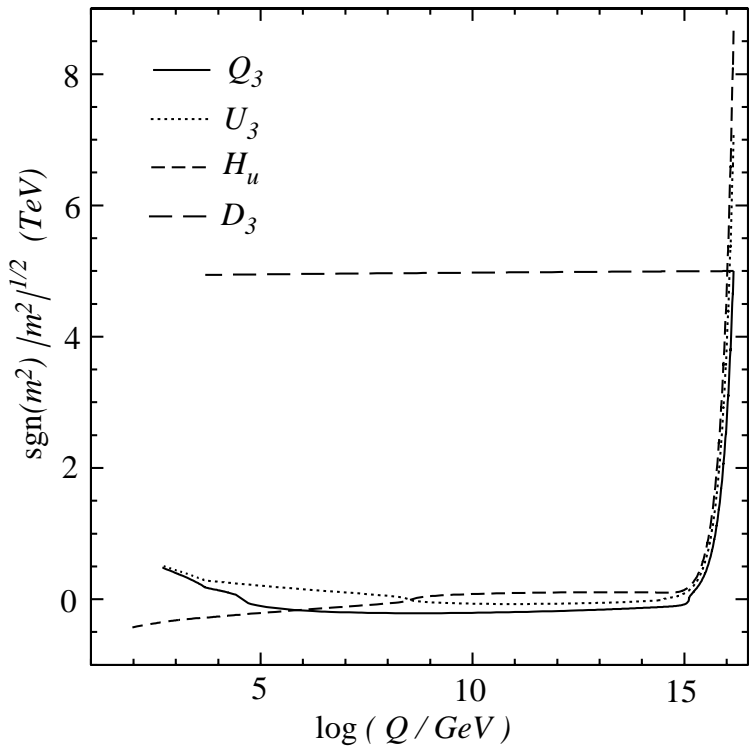}{0.65}
\caption{The two-loop renormalization group evolution of SSB masses
for a representative case in the low $\tan\beta$ QFP scenario.  The
boundary conditions at $Q \simeq 2.4 \times 10^{16}$ GeV are those of
Eq.~(\ref{bc}) with $m_0 = 5$ TeV, and $M_{1/2} = 500$ GeV and $A_0 =
0$.  The SSB masses for $Q_3$, $U_3$, and $H_u$ are quickly driven to
the weak scale by the top Yukawa coupling, while the rest of the
scalars, represented by $D_3$ here, remain at the multi-TeV scale.}
\label{fig:fig1}
\end{figure}

\begin{figure}[ht]
\postscript{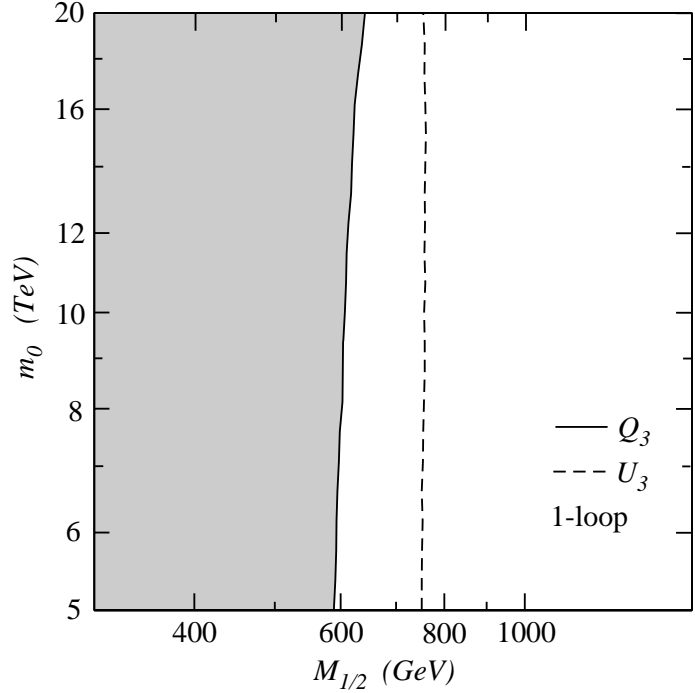}{0.60}
\caption{The allowed region (shaded) in the $(M_{1/2}, m_0)$ parameter
place, where $M_{1/2}$ is the high scale gaugino mass, and $m_0$
specifies the high scale scalar masses through Eq.~(\ref{bc}).
One-loop renormalization group equations are used and $A_0=0$.  To the
left of the solid (dashed) contour, the scalar $Q_3$ ($U_3$) mass is
below 1 TeV at the weak scale.  All physical squared masses are
positive and electroweak symmetry is properly broken throughout the
plane. Note the different mass scales.}
\label{fig:fig2}
\end{figure}

\begin{figure}[ht]
\postscript{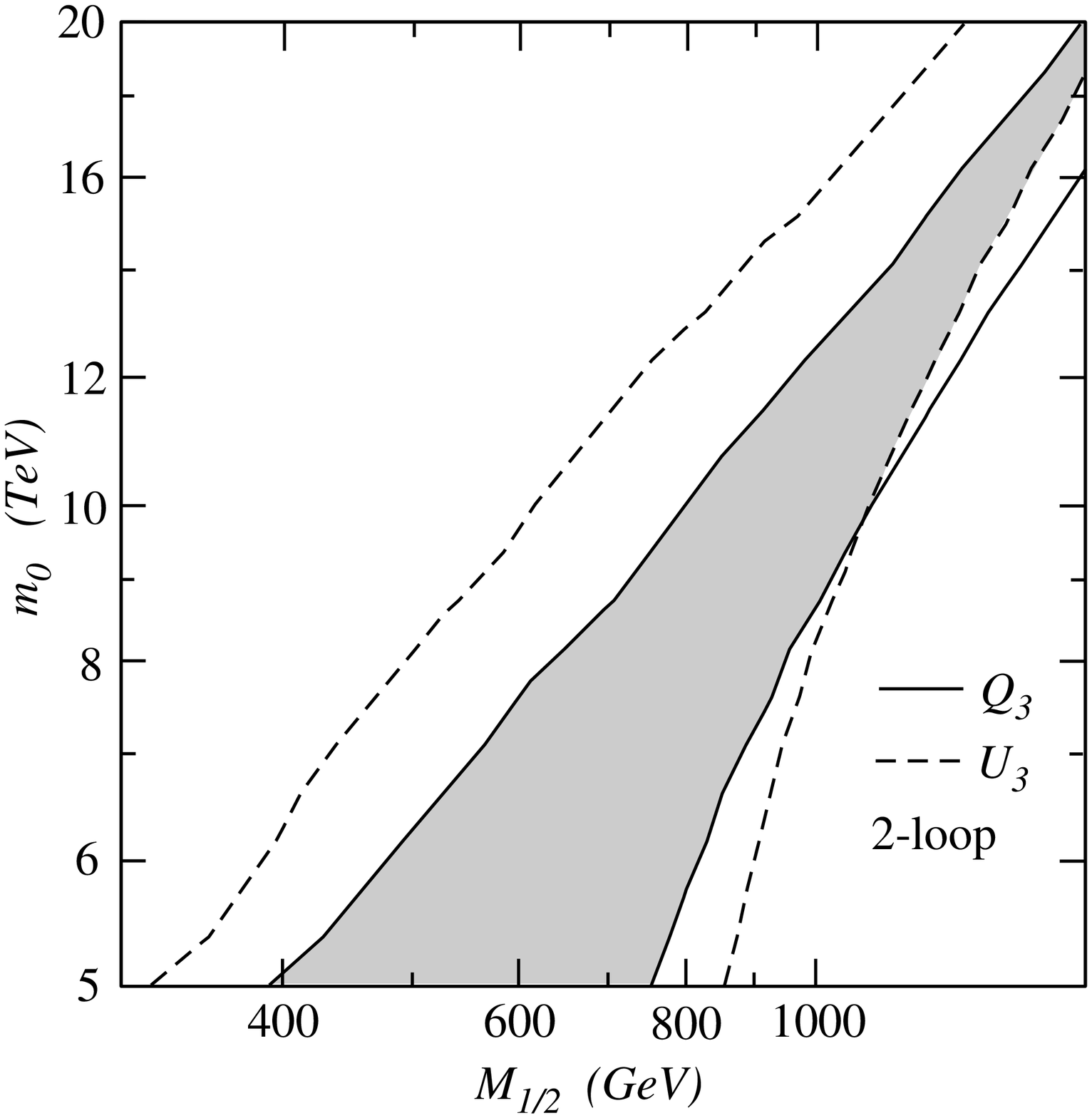}{0.60}
\caption{Same as in Fig.~\ref{fig:fig2}, but including the two-loop
gauge contributions in the evolution.  To the left of the left-most
solid (dashed) contour, the scalar $Q_3$ ($U_3$) mass is negative at
the weak scale; to the right of the right-most solid (dashed) contour,
the scalar $Q_3$ ($U_3$) mass is above 1 TeV.  In the shaded region,
both masses are positive and below 1 TeV.}
\label{fig:fig3}
\end{figure}

We have confirmed the analytic approximations described above with
complete numerical calculations including the two-loop gauge
corrections.  In Fig.~\ref{fig:fig1} we show the renormalization group
evolution of the SSB mass parameters for a representative set of
boundary conditions satisfying Eq.~(\ref{bc}).  Despite multi-TeV
values at the high scale boundary, we see that the masses renormalized
by the large top Yukawa are quickly driven to the weak scale in the
infra-red.  The other scalar masses remain at the multi-TeV scale, and
we see that a scalar mass hierarchy is generated radiatively.
To quantify how generic such results are, we display in
Figs.~\ref{fig:fig2} and \ref{fig:fig3} the regions in parameter space
for which phenomenologically-desirable squark masses are obtained. In
Fig.~\ref{fig:fig2} the weak scale parameters are obtained from the
high scale boundary conditions through one-loop renormalization group
equations.  In the shaded region, both $Q_3$ and $U_3$ masses are
positive and below 1 TeV.  Any gaugino mass is possible, as long as it
is not so large as to drive the $Q_3$ and $U_3$ masses above 1 TeV.
In Fig.~\ref{fig:fig3}, the two-loop gauge contributions are
included. As noted above, these contributions must be compensated by
gaugino contributions to avoid tachyons and color-breaking minima, and
so now, for a given $m_0$, there is a minimum allowed $M_{1/2}$.  We
see, however, that there is still a substantial band in which all
phenomenological requirements are met, and the $Q_3$ and $U_3$ masses
are below a TeV.

In this scenario, the fine-tuning associated with $m_{H_u}$ and the
squark fields $Q_3$ and $U_3$, which are strongly coupled to $H_u$,
has been successfully eliminated.  Unfortunately, $m_{H_d}$ is not
affected by the Yukawa fixed point in the low $\tan\beta$ scenario and
remains at the heavy scale.  Electroweak symmetry breaking therefore
requires $\mu^2 \sim m_{H_d}^2 \sim \mheavy^2$, and this scenario is
still fine-tuned.\footnote{The requirement that squark mixing not lead
to color-breaking minima also leads to the constraint $\mlight m_{H_d}
< m_{Q}^{2} \simeq 7M_{1/2}^{2}$, which is, however, weaker than the
constraints discussed above and is easily satisfied.}

This flaw may be avoided in high $\tan\beta$ scenarios, to which we
will turn in the following section.  Before doing so, however, we
collect here a number of remarks.  First, note that the boundary
conditions of Eq.~(\ref{bc}) are inconsistent with any minimal GUT
embedding requiring $m_{Q_3}(0) = m_{U_3}(0)$. Second, large ($\agt
10\%$) and negative finite mass corrections to the top quark mass may
increase the low $\tan\beta$ QFP value to $\tan\beta^{\rm FP} \gg
1$. In this case, as is evident from Eq.~(\ref{finetune}), the
fine-tuning related to large $m_{H_d}$ is significantly diminished,
and $m_{H_d} \sim \mheavy$ may be tolerated.  However the finite mass
contributions realized by supersymmetric QCD corrections in most
models are $\alt 10\%$, and so this scenario may be difficult to
realize.

Finally, it is entertaining to note that $m_{H_d}, |\mu| \gg m_{\rm
weak}$ is actually preferred by coupling constant unification, as it
leads to a pattern of superparticle threshold corrections that
diminishes the prediction for the strong coupling $\alpha_{s}(m_{\rm
weak})$.  In the absence of threshold corrections, one predicts
too-large $\alpha_{s}(M_{Z}) \approx 0.13$, and most typical patterns
of superparticle threshold corrections only aggravate this problem.

\section{Radiative Hierarchy with High $\tan\beta$}

The fine-tuning situation may be resolved in the case of high
$\tan\beta \sim 50 - 60$, where both $h_t$ and $h_b\equiv h_d^{33}$
are near their fixed points.  The coupled set of renormalization group
equations is now more complicated.  However, assuming $h_t \approx
h_b$, and neglecting for simplicity all gaugino masses, trilinear
scalar couplings and Tr$[Ym^2]$ as before, we find that the scalar
masses evolve as
\begin{equation}
\frac{d\,{\bf  m^2}}{d t} = \frac{h_t^2}{4\pi} 
{\bf X}_{\rm high} {\bf m^2} \ ,
\end{equation}
where
\begin{equation}
{\bf X}_{\rm high} = - \left(
\begin{array}{ccccc}
3 & 3 & 3 & 0 & 0 \\
2 & 2 & 2 & 0 & 0 \\
1 & 1 & 2 & 1 & 1 \\
0 & 0 & 2 & 2 & 2 \\
0 & 0 & 3 & 3 & 3 \end{array} \right)
\end{equation}
and here ${\bf m^2} = (m_{H_u}^2, m_{U_3}^2, m_{Q_3}^2, m_{D_3}^2,
m_{H_d}^2)^T$.  Three eigenvectors of ${\bf X}_{\rm high}$ have
eigenvalue 0; the other two are ${\bf\hat{m}_1^2} = (-3,-2,0,2,3)^T$
with eigenvalue $-5$ and ${\bf\hat{m}_2^2} = (3,2,2,2,3)^T$ with
eigenvalue $-7$.  We thus expect a two parameter family of boundary
conditions leading to a scalar mass hierarchy.

As before, we consider the QFP framework, but now for high
$\tan\beta$.  Neglecting gaugino masses and the trilinear scalar
couplings, we find simple solutions for the low-energy masses in terms
of their high scale boundary values~\cite{solution2}:
\begin{eqnarray}
m_{H_u}^2 &\simeq& m_{H_u}^{2}(0) - 3 \Delta m_t^2 \nonumber \\
m_{H_d}^2 &\simeq& m_{H_d}^{2}(0) - 3 \Delta m_b^2 \nonumber \\
m_{Q_i}^2 &\simeq& m_{Q_i}^{2}(0) - \delta_i
  (\Delta m_t^2 + \Delta m_b^2) \nonumber \\
m_{U_i}^2 &\simeq& m_{U_i}^{2}(0) - \delta_i 2
  \Delta m_t^2 \label{hightanbeta} \\
m_{D_i}^2 &\simeq& m_{D_i}^{2}(0) - \delta_i 2 \Delta m_b^2
\nonumber \\
m_{L_i}^2 &\simeq& m_{L_i}^{2}(0) \nonumber \\
m_{E_i}^2 &\simeq& m_{E_i}^{2}(0) \nonumber \ ,
\end{eqnarray}
where
\begin{eqnarray}
\Delta m_t^2 &\simeq& \frac{1}{7} \left[ m_{H_u}^{2}(0) +
m_{Q_3}^{2}(0) + m_{U_3}^{2}(0) \right] r \nonumber \\
 &-&\frac{1}{10} \left[ m_{H_u}^{2}(0) - m_{H_d}^{2}(0) +
m_{U_3}^{2}(0) - m_{D_3}^{2}(0) \right] \left[ \frac{5}{7} r +
(1-r)^{\frac{5}{7}} - 1 \right] \\
\Delta m_b^2 &\simeq& \frac{1}{7} \left[ m_{H_d}^{2}(0) +
m_{Q_3}^{2}(0) + m_{D_3}^{2}(0) \right] r \nonumber \\
 &+&\frac{1}{10} \left[ m_{H_u}^{2}(0) - m_{H_d}^{2}(0) +
m_{U_3}^{2}(0) - m_{D_3}^{2}(0) \right] \left[ \frac{5}{7} r +
(1-r)^{\frac{5}{7}} - 1 \right] \ ,
\end{eqnarray}
and, as before, $r = \left[ h_t/h_t^{\rm FP} \right] ^2 \leq 1$.  In
this solution, we have neglected small differences in the top and
bottom Yukawa coupling evolution, and assumed vanishing leptonic
couplings.  In particular, we neglect $h_{\tau}$; see the discussion
below.

In contrast to the previous low $\tan\beta$ case, $h_b$ is now
significant.  We must then demand that the $H_u$, $H_d$, $Q_3$, $U_3$,
and $D_3$ scalar masses all be driven to zero in the infra-red.  We
find that this scenario is obtained for an extremely simple
two-parameter family of boundary conditions given by
\begin{eqnarray}
m_{H_u}^2 (0) &=& \frac{3}{2} m_{U_3}^{2}(0) \nonumber \\
m_{H_d}^2 (0) &=& \frac{3}{2} m_{D_3}^{2}(0) \label{bc2} \\
m_{Q_3}^2 (0) &=& \frac{1}{2} \left[ m_{U_3}^{2}(0) +
m_{D_3}^{2}(0) \right] \nonumber \ .
\end{eqnarray}
These boundary conditions are just a reparametrization of the space
spanned by the eigenvectors ${\bf\hat{m}_1^2}$ and ${\bf\hat{m}_2^2}$.
Clearly, in this case both Higgs mass parameters are affected by the
fixed point and no fine-tuning is required (aside from the moderate
tuning at the level of $m_W^2/\mlight^2$, which is always associated
with such high values of $\tan\beta$~\cite{largetgb}).

The solutions of Eq.~(\ref{hightanbeta}) are valid for boundary values
$h_{\tau}^{2}(0) \ll h_{b}^{2}(0)$ and $h_{\tau}\ll 4\pi\,\mlight
/\mheavy$.  The first relation is found in a certain range of very
high $\tan\beta \sim 50-60$.\footnote{Again, the exact value of
$\tan\beta$ for which $h_{t}(0) \simeq h_{b}(0) \gg h_{\tau}(0)$
depends sensitively on low-energy finite radiative corrections to the
$t$ and $b$-quark masses and on the exact value of the strong
coupling.}  (See, for example, Fig.~1 of Ref.~\cite{fixed}.)  One
often associates the high $\tan\beta$ QFP scenario with either $h_{b}
= h_{\tau}$ or $h_{t}= h_{b} = h_{\tau}$ unification at the GUT scale,
as implied by minimal SU(5) and SO(10) GUTs, respectively.  A subset
of the solutions of Eq.(\ref{bc2}), with $m_{Q_3}^2(0) = m_{U_3}^2(0)
= m_{D_3}^2(0)$ and $m_{H_u}^{2}(0) = m_{H_d}^{2}(0) = \frac{3}{2}
m_{Q}^{2}(0)$, is consistent with such a GUT embedding.  (The
hypercharge trace condition is automatically satisfied in this case.)
In general, however, the boundary conditions need not admit a true
(minimal) GUT embedding, and we therefore do not require such scalar
mass relations or the accompanying Yukawa coupling unifications.

It is difficult to incorporate analytically the effects of a
non-negligible $h_{\tau}$, and generally an involved numerical
analysis is required.  In Ref.~\cite{solution2} this effect was
estimated, but the results were valid only for $0.6 \alt r \alt
0.95$. The required boundary conditions have a complicated dependence
on $r$ and therefore do not have obviously simple forms away from the
QFP value of $r=1$.

In the quantitative discussions above, we have focused on only two
simple scenarios with minimal field content.  It should be stressed,
however, that while the required boundary conditions depend on the
specific Yukawa fixed point structure, the existence of such boundary
conditions stems from the general existence of such a structure.
Hence, it is reasonable to speculate that our observations apply more
generally.  For example, one could look for similar QFP solutions in
the MSSM extended by a gauge singlet $S$ interacting through the
superpotential term $SH_u H_d$, or at the case of lepton number
violating Yukawa couplings with simultaneous fixed points~\cite{lnv}.
Many other such examples are possible.

\section{High-energy Frameworks and $R$ Symmetry}
\label{sec:Rsymm}

We have seen that inverted scalar hierarchies may be generated
radiatively for certain boundary conditions.  Such boundary conditions
are both highly constrained and highly predictive, and it is of some
interest to investigate specific high energy frameworks that give such
mass patterns.  Here we will limit ourselves to a discussion of
general principles that lead to the required features.

Let us concentrate on the high $\tan\beta$ scenario.  The appearance
of a light-heavy hierarchy in the scalar mass sector can only occur if
there is already a hierarchy between the scalar masses (heavy) and the
$\mu$ parameter, $m_3^2$, gaugino masses, and $A$-terms (light).
(Electroweak symmetry breaking requires $m_3^2$ at the light scale,
since $m_3^2 = \frac{1}{2}(m_{H_u}^2 + m_{H_d}^2 + 2\mu^2)
\sin2\beta$.)

Such a hierarchy might be generated by an approximate U(1) symmetry.
In the absence of the $\mu$ and SSB parameters, the MSSM possesses two
global U(1) symmetries: a Peccei-Quinn symmetry, under which all
components of a given superfield have the same charge, and an $R$
symmetry, under which the boson and fermion components of a given
superfield differ by one unit. If $\mu$ and the SSB parameters are
viewed as spurion fields~\cite{dt}, we may choose the following charge
assignments for them: $PQ(\mu) = PQ(m_3^2) = 1$, with all others PQ
neutral, $R(m_0) = R(\mu) = 0$, and $R(m_3^2) = R(M_{1/2}) = R(A) =
-2$.  Under suitable linear combinations of these two symmetries, such
as $R + PQ$, all parameters are charged, except for the scalar masses
$m_0$. (Note that scalar masses are neutral, and gaugino masses and
$A$-terms are charged, for all possible linear combinations.)  Thus,
an approximate U(1) symmetry, such as an $R+PQ$ symmetry, naturally
produces the necessary hierarchy, and the presence of $R$ symmetries
may play a vital role in realizing models that exhibit the inverted
hierarchy.

Alternatively, the suppression of the necessary parameters may be the
result of some other mechanism.  Assume, for example, that the scale
at which SUSY-breaking is communicated to the MSSM, $M$, is
significantly higher than the initial scale of SUSY-breaking itself,
$\sqrt{F}$.  We can then express the most general set of operators in
an expansion in powers of $\sqrt{F}/M$. The leading terms in that
expansion that generate the $\mu$ parameter and soft terms have the
following form:
\begin{eqnarray}
{\rm Scalar \ masses:} && \int d^4 \theta\, \Phi_i^\dagger \Phi_i
\left[ \frac{S^\dagger S}{M^2} + \frac{Z^\dagger Z}{M^2} +
\cdots \right] \label{scalar}\\
\mu \ {\rm parameter:} && \int d^4 \theta\, H_u H_d
\left[ \frac{S^\dagger}{M} + \cdots \right] \label{mu}\\
{\rm Gaugino \ masses:} && \int d^2\theta\, W^{\alpha}
W_{\alpha} \left[ \frac{S}{M} + \cdots \right] \label{gaugino}\\
A{\rm -terms:} && \int d^2 \theta\, \Phi_i \Phi_j \Phi_k
\left[ \frac{S}{M} + \cdots \right] \ {\rm and}
\int d^4 \theta\, \Phi_i^{\dagger} \Phi_i \left[ \frac{S}{M} +
\cdots \right] \label{Aterm}\ ,
\end{eqnarray}
where the $W_{\alpha}$ are gauge vector supermultiplets containing the
standard model gauginos, the $\Phi_i$ are standard model chiral
superfields, and $S$ and $Z$ represent SUSY-breaking gauge singlet and
non-singlet superfields, respectively. These terms give SSB parameters
and the $\mu$ parameter when the $S$ and $Z$ fields get $F$-term
vacuum expectation values: $S \to F_S \theta^2$, $Z \to F_{Z}
\theta^2$ (and, in the second source for $A$-terms, $\Phi^{\dagger}_i
\to F^*_{\Phi_i} \bar{\theta}^2 \sim \Phi_j \Phi_k\bar{\theta}^2$).

{}From the expressions above, it is clear that the terms corresponding
to dimension 3 operators rely on SUSY-breaking singlet fields at
leading order in $M^{-1}$, while the scalar masses do not.  Therefore,
in any scenario in which $F_S \ll F_Z$ (or $S$ is absent from the
spectrum), $\mu$, the gaugino masses, and all $A$-terms will be
suppressed relative to scalar masses.  For example, to generate the
desired hierarchy, it is sufficient for $F_S$ to be generated
radiatively so that $F_S \sim \alpha/4\pi \, F_Z$. (Note, however,
that $m_3^2$ must be suppressed by some other means, such as the U(1)
symmetries discussed above.)

Finally, it is interesting to ask whether such a hierarchy could ever
occur in supergravity-mediated SUSY-breaking models. It is known that
in models without singlets, gaugino masses are suppressed relative to
scalar masses. If we further assume that there are no Planck scale
vacuum expectation values in the hidden sector (as is expected in
models that break SUSY in the flat limit), then $A$-terms will also be
suppressed~\cite{joichi}.  In such scenarios, conventional
contributions to the gaugino masses and $A$-terms are highly
suppressed, and the dominant contributions have recently been shown to
be those arising from the superconformal anomaly~\cite{anomaly}. In
fact, the natural suppression of gaugino masses relative to squark
masses is then one-loop, roughly corresponding to the size we require
in our mechanism.

Once the hierarchy between the scalar and gaugino masses is generated,
it is still necessary to understand the particular form of the scalar
mass boundary conditions that are required in these scenarios.

The rational relations that are required among the soft masses in the
previous sections are immediately reminiscent of the relations one
would expect were soft masses to be communicated via $D$-terms of
broken gauge symmetries.  This results in terms ${\cal L} =
\frac{g^2}{2} [\tr\,Q_i m_i^2\pm\xi]^2$, where $Q$ is a charge in the
Cartan subalgebra of the broken group and $\xi$ is an order parameter
of the group's breaking.  Note that the squared masses are always
proportional to the broken Cartan charges of the
fields. Unfortunately, it is impossible to have the spectrum of
charges corresponding to Eq.~(\ref{bc}) [Eq.~(\ref{bc2})] and
simultaneously demand invariance of the top [top and bottom] Yukawa
coupling under the broken symmetry group, which is a natural
assumption given its large size.

One might also consider the framework of weakly coupled string theory.
There, the dilaton field is one of the singlets $S$ above.  If
SUSY-breaking is dominated by the dilaton $F$ component, then gaugino
masses and (universal) scalar masses are of the same order~\cite{ln}.
However, if SUSY-breaking is dominated by $F$ components of moduli
$\Phi_M$, gaugino masses arise only at loop-level in string theory,
giving $M_{1/2} \sim (\alpha_{\rm string}/4\pi)\, m_{0}$. (Of course,
a mechanism for suppressing the $A$-terms is also needed.)
Incidentally, in the moduli-dominated scenario, scalar masses are
generically all at the same scale, but may differ by order one
coefficients given by the K\"ahler metric: $m_{i}^{2} \approx K_{ii}
m_{0}^{2}$.  This is exactly the necessary condition for radiative
inverted hierarchy generation.  In this framework, the boundary
condition scalar mass ratios that we derived above correspond to
ratios of modular weights of the different fields.

\section{The $CP$ and Polonyi Problems}
\label{sec:CP}

The $R$ symmetry discussed above was previously studied in
Ref.~\cite{ftp}, where a number of attractive phenomenological
features were noted.  In that work, an approximate $R$ symmetry was
seen as a possible source for a hierarchy $\mu, M_{1/2}, A \sim
1\,{\rm GeV} \ll m_{0} \sim m_{\rm weak}$.  In this study, we are
considering mass scales roughly 100 times those discussed there.
However, as most of the attractive features discussed there result
from the hierarchy itself, they apply equally well here. For example,
supersymmetric contributions to electron and neutron electric dipole
moments are $d_{e,n} \propto (1/m_0^2 ) ( M_{1/2} \tilde{m} / m_0^2
)$, where $\tilde{m} \sim \mu, A$, and $m_0$ represents scalar masses
of the first generation.  These contributions are therefore suppressed
both by the large scalar mass scale and by the hierarchy between
$\mlight$ and $\mheavy$, and are well below current experimental
bounds.

Our models also have an important cosmological virtue.  Many
supergravity models contain a boson $\phi$, the Polonyi field, with
mass of order the gravitino mass.  The Polonyi field has gravitational
couplings and, consequently, an extremely long lifetime $\tau \sim
M_P^2/m_{\phi}^3$, where $M_P$ is the Planck mass.  For such models
with gravitino masses of order 100 GeV, the Polonyi field typically
decays during or after temperatures of order 1 MeV, thereby
potentially ruining nucleosynthesis.  This is often referred to as the
``Polonyi problem''~\cite{Polonyi}, and is a serious cosmological
difficulty for many models.

The Polonyi problem may be solved, for example, in particular
SUSY-breaking scenarios~\cite{Polsolns}. Irrespective of the
SUSY-breaking mechanism, however, in the models discussed here, the
Polonyi problem is always alleviated, as the gravitino mass $m_{3/2}
\sim m_{0}$ is in the multi-TeV range.  It has been pointed out that
this provides a solution to the Polonyi problem, since in this case
even a Polonyi field with mass $m_{\phi} \sim m_{3/2} \sim 10$ TeV
decays sufficiently quickly to avoid the difficulty mentioned
above~\cite{ENQ}. Potential problems with generating the baryon
asymmetry and overclosing the universe with Polonyi decay products may
also be solved, the first with Affleck-Dine baryogenesis, and the
second with the presence of a very light and stable superpartner or
with $R$-parity violation~\cite{moroi}.

\section{Summary and Outlook}
\label{sec:summary}

To conclude, we have investigated the possibility that soft
SUSY-breaking scalar mass parameters are not $\alt 1$ TeV at some high
scale boundary, as is typically assumed, but rather, are all in the
multi-TeV range.  For particular boundary conditions, given in
Eqs.~(\ref{bc}) and (\ref{bc2}), we find that scalars with large Higgs
couplings are asymptotically driven to the weak scale by
renormalization group evolution, while the remaining scalars stay at
the multi-TeV scale.  By this mechanism, the light scalars are
precisely those that must be light to preserve the gauge hierarchy,
and the heavy scalars are precisely those corresponding to light
fermions that must be heavy to satisfy stringent flavor-changing
constraints.

As in all models with hierarchical squark masses, it is important to
note that multi-TeV scalar masses by themselves do not completely
satisfy all flavor constraints~\cite{am,ag}.  In the above analysis,
we have discussed only the evolution of the flavor diagonal masses.
However, it is possible that off-diagonal masses are present at the
high scale; such masses are largely unaffected by renormalization
group evolution.  Recent improvements in calculations of
$K^0-\bar{K}^0$ mixing have strengthened this most stringent
constraint, so that now, even with $\mheavy \sim 10$ TeV, the
off-diagonal squark masses must roughly satisfy $m_{12}^2 / \mheavy^2
\alt 0.1$~\cite{constraints}.  This requirement on mixings (or
non-degeneracies) is, however, relatively mild and is a great
improvement over analogous constraints on models with squarks below
the TeV scale.

There are several experimental signatures of these models.  As evident
{}from the discussion above, detectable effects in the kaon system are
possible.  In addition, although the requirement of no tachyons
implies roughly $m_{13}^2, m_{23}^2 \alt \mlight\mheavy$, large
effects in the $B$ system, for example, may be possible, and are
potentially observable at current of near-future
experiments~\cite{bphys}.  There are also implications for the high
energy frontier.  At least some gauginos and some third generation
sfermions are predicted to be accessible at the next generation of
collider experiments.  While the $\mheavy$ sector will not be, it may
then be explored indirectly by measurements of the superoblique
corrections of Ref.~\cite{so}.  Although very massive scalars decouple
from many observables, they leave their imprint on low energy
processes by breaking the equality of gauge boson-fermion-fermion
couplings and the corresponding gaugino-fermion-sfermion couplings.
These deviations are non-decoupling.  The superoblique parameters are
therefore sensitive to arbitrarily heavy MSSM sfermions, and may be
measured to high accuracy in processes involving the observable
superparticles~\cite{so,kn}.

In this scenario, several requirements must be met.  First, the flavor
off-diagonal masses discussed above must be suppressed relative to
flavor diagonal ones.  Of course, as noted above, the necessary
suppressions are mild relative to models with all scalars below the
TeV scale.  It is also worth noting that in such conventional models,
even if some mechanism for suppressing flavor violation is
implemented, (flavor-conserving) constraints on electric dipole
moments and the Polonyi problem may still be rather severe; as argued
in Sec.~\ref{sec:CP}, these problems are naturally alleviated in the
models discussed here.

In addition, the requirement of extreme scalar degeneracy or alignment
to remove dangerous flavor-changing contributions is replaced by the
requirement of particular high scale boundary conditions.  In the
absence of a more fundamental theory, this is not an obvious
improvement.  However, this scenario opens a new arena for SUSY model
building.  With regard to the supersymmetric flavor problem, it
presents the possibility that a solution is provided by some dynamical
mechanism that produces the required boundary conditions, such as the
simple conditions of Eq.~(\ref{bc2}).  We have discussed theoretical
motivations for the required hierarchies and a possible relation to
$R$ symmetries.  More generally, and independent of the SUSY flavor
problem, it raises the possibility of scenarios in which electroweak
symmetry breaking is {\em not} fine-tuned, even though the fundamental
scale for the soft SUSY-breaking parameters is $\sim 10$ TeV, rather
than $\sim 1$ TeV as is typically assumed.

Finally, we note that, while our illustrations have been limited to
the MSSM, these observations should apply more generally.  It would be
particularly interesting to pursue this framework in models with
extended fixed point structures, and also more extensively in the high
$\tan\beta$ regime.

\section*{Acknowledgments}

The authors thank J.~Bagger, P.~G.~O.~Freund, T.~Moroi and A.~Pomarol
for stimulating comments and conversations.  JLF is grateful to the
theory groups of Stanford and SLAC for hospitality, and NP thanks the
theory group at CERN for its hospitality.  JLF is supported by the
Department of Energy under contract DE--FG02--90ER40542 and through
the generosity of Frank and Peggy Taplin.  CK is supported by the
Department of Energy under contract DE--AC03--76SF00098.  The work of
NP is supported by the NSF under grant NSF--PHY--94--23002 and by the
Department of Energy under contract DE--FG02--96ER40559.

\end{document}